# Robust evaluation of vaccine effects based on estimation of vaccine efficacy curve

Ziwei Zhao[1], Xiangmei Ma[1], Paul Milligan[2], Yin Bun Cheung[1,3,4]

1. Centre for Quantitative Medicine, Duke-NUS Medical School, 20 College Road, Singapore 169856

2. Faculty of Epidemiology and Population Health, London School of Hygiene & Tropical Medicine, Keppel Street, London WC1E 7HT, UK

3. Signature Programme in Health Services & Systems Research, Duke-NUS Medical School, 8 College Road, Singapore 169857

4. Tampere Center for Child, Adolescent and Maternal Health Research, Tampere University, Arvo Ylpön katu 34, Tampere 33520, Finland

Correspondence:

Ziwei Zhao, Centre for Quantitative Medicine, Duke-NUS Medical School, Singapore, 20 College Road, Singapore 169856

zhaoziwei@u.duke.nus.edu

Tel: +65 97746293

Word counts: 3667

## Abstract

**Background**

The Cox model and its extensions assuming proportional hazards is widely used to estimate vaccine efficacy (VE). In the typical situation that VE wanes over time, the VE estimates are not only sensitive to study duration and timing of vaccine delivery in relation to disease seasonality but also biased in the presence of sample attrition. Furthermore, estimates of vaccine impact such as number of cases averted (NCA) are sensitive to background disease incidence and timing of vaccine delivery. Comparison of the estimates between trials with different features can be misleading.

**Methods**

We propose estimation of VE as a function of time in the Cox model framework, using the area under the VE curve as a summary measure of VE, and extension of the method to estimate vaccine impact. We use simulations and re-analysis of a RTS,S/AS01 malaria vaccine trial dataset to demonstrate their properties and applications.

**Results**

Simulation under scenarios with different trial duration, magnitude of sample attrition and timing of vaccine delivery, all assuming vaccine protection wanes over time, demonstrated the problems of conventional methods assuming proportional hazard, robustness and unbiasedness of the proposed methods, and comparability of the proposed estimates of vaccine efficacy and impact across trials with different features. Furthermore, the proposed NCA estimators are informative in determining the optimal vaccine delivery strategy in regions with highly seasonal disease transmission.

**Conclusions**

The proposed method based on estimation of vaccine efficacy trajectory provides a robust, unbiased, and flexible approach to evaluate vaccine effects.

Key messages:

- Estimation of vaccine efficacy based on proportional hazard models is sensitive to trial implementation features and is biased in the presence of substantial sample attrition when vaccine efficacy wanes over time, and it does not facilitate fair comparison between trials.

- Conventional estimators of vaccine impact are sensitive to background disease incidence and seasonality and are difficult to compare between trials.

- We proposed estimation of vaccine efficacy as a function of time, using area under the vaccine efficacy curve as a summary measure of efficacy, and an extension of the method to evaluate number of cases averted and number needed to vaccinate to prevent one case.

- The proposed methods are robust to variation in trial features, and they can facilitate more reliable comparison between clinical trials and provide insights in vaccine delivery strategy options.

# Introduction

In studies of infectious disease prevention, vaccine efficacy (VE) is often defined as one minus hazard ratio (HR), where HR is estimated from randomized controlled trials using the Cox model or its extensions, such as the Andersen-Gill model[1-3]. We refer to them collectively as Cox-type models. The estimation often assumes proportional hazards (PH), i.e. HR and VE are assumed constant over time. However, protective interventions against infectious diseases tend to have waning efficacy over time, such as malaria, rotavirus and COVID-19 vaccines, which violates the PH assumption[4-6].

While a "wrong model" based on PH assumption may sometimes provide useful information about intervention effect averaged over time[7-8], in the vaccine context it misses out important information that is needed for clinical and policy decision making, such as the timing VE may wane to a level that mandates a booster dose. One consequence of making the PH assumption when in fact VE wanes over time is the sensitivity of VE estimates to trial duration. Some less known consequences include over-estimation of VE in the presence of sample attrition and sensitivity to timing of vaccination delivery for diseases with strong seasonality. These factors and variations in background disease incidence also affect number of cases averted (NCA), which is an important measure of vaccine impact[4,8-9]. These issues not only make VE estimates based on PH assumption difficult to interpret but also render comparison of results from different trials error-prone. For example, Rosenthal compared the efficacy of the WHO recommended RTS,S/AS01 malaria vaccine and the seemingly higher efficacy of R21 malaria vaccine[10]. However, Birkett et al. argued that this comparison is unreliable due to difference in timing of vaccine delivery in the presence of strong seasonality in disease incidence[11]. There were also concerns about interpretation and comparison of Covid-19 VE estimates including variations in trial duration and background disease incidence[12]. Such debates emphasized the importance of developing analytic methods that can facilitate robust estimation and reliable comparisons.

Some studies have considered capturing time-varying intervention effect in infectious disease prevention trials by incorporating parametric functions[13-15], step functions[16-17], or splines[18-20] into Cox-type models. They focused on understanding how the protective efficacy of a medicinal product changed over time, as opposed to providing summary measures and enabling valid comparison of trial results.

In this article, we demonstrate the bias and non-comparability of VE estimates from Cox-type models based on the PH assumption. Then we propose estimation of VE as a smooth function of time. We introduce the concept of area under VE curve (AUC) as a summary measure of vaccine efficacy that is not sensitive to trial implementation features. We will illustrate flexible applications of the method to estimate NCA under different delivery strategies and timing of vaccination in regions with strong seasonal transmission and different levels of disease incidence. We will evaluate the properties of models based on PH assumption versus our proposed methods by simulation, and demonstrate their application in re-analysis of a RTS,S/AS01 malaria vaccine trial dataset.

## Development

**Cox-type models with proportional hazard assumption**

The basic form of Cox proportional hazard model assumes:

$$\lambda_i(t) = \lambda_0(t) exp(\beta z_i),$$

where $\lambda_0(t)$ is an unspecified baseline hazard function, $z_i$ is the intervention status of subject $i$, with $z_i = 1$ denoting vaccine and $z_i = 0$ denoting control/placebo, and $\beta$ is the ln(HR) to be estimated. The VE estimate, $\widehat{VE} = 1 - exp(\hat{\beta})$, where $\hat{\beta}$ is the estimated ln(HR), is time-constant. Let $t_{(1)} < t_{(2)} \cdots < t_{(N)}$ be the ordered event times, $z_j$ be the intervention status of the person who experiences the outcome event at $t_{(j)}$, and $R(t_{(j)})$ be the set of persons who are at risk just before time $t_{(j)}$. The ln(HR) is estimated by maximizing the partial likelihood

$$PL(\beta) = \prod_{j=1}^{N} \frac{exp(\beta z_j)}{\sum_{k \in R(t_{(j)})} exp(\beta z_k)}. \qquad (1)$$

The AG and Cox model have the same formulation of partial likelihood as in equation (1). However, in the Cox model, a person who has experienced his/her first event at $t_{(j-1)}$ or earlier is excluded from $R(t_{(j)})$ or later risk sets. In the AG model there is no such exclusion. Previous studies have shown that the estimation based on time-to-first-event suffers a selection bias[7,21,22], while the AG model is robust to this bias[3,23]. The AG model is the method of choice if the outcome event is repeatable.

Note that the partial likelihood in equation (1) consists of contributions from N observed event times. When the PH assumption is violated, $\hat{\beta}$ may be interpreted as an average of the underlying time-varying effect, $\beta(t)$, over the range of event times observed in the data[7,24]. Therefore, different study designs and implementation that affect the distribution of the observed event times will lead to different composition of the partial likelihood and, in turn, different $\widehat{VE}$. Other factors being the same, trials with shorter follow-up or heavier sample attrition (censoring) have observed event times more concentrated in the earlier part of post-vaccination time. In realistic situations that VE monotonically declines over time, $\widehat{VE}$ estimated from these trials would tend to be larger.

[Figure 1 about here]

$\widehat{VE}$ is also affected by timing of vaccine delivery. Consider a 12-month trial that starts at the beginning of a 6-month low disease transmission season and the true $VE(t)$ has a wanning trajectory as shown in Figure 1a. If there is no sample attrition, the majority of the events will be observed during the latter half of the study time, when disease incidence quadruples and the true $VE(t)$ declines rapidly. Events observed during the period in which $VE(t)$ is lower will dominate the composition of the partial likelihood, and $\widehat{VE}$ will be small. Conversely, $\widehat{VE}$ would be larger if the trial starts at the beginning of the high disease transmission season (Figure 1b).

## Cox-type models with time-varying effects

It is possible to allow time-varying effects in Cox-type models. The hazard function for the $i$-th person is adapted to:

$$\lambda_i(t) = \lambda_0(t)exp[f(t)z_i], \qquad (2)$$

where $f(t)$ is a function of time. The time-varying HR is given by $HR(t) = exp[f(t)z_i]$. There are various choices for $f(t)$[14,15]. In per protocol analysis of vaccine trials, analysis time usually starts at 14 days after completion of primary series. In this manuscript, $t = 0$ refers to this starting point. In this context $f(t)$ likely declines monotonically[15]. For purpose of demonstration, here we consider $f(t) = \beta_0 + \beta_1 t$ as an example. Equation (2) can be expanded to:

$$\lambda_i(t) = \lambda_0(t)exp[(\beta_0 + \beta_1 t)z_i] = \lambda_0(t)exp(\beta_0 z_i + \beta_1 tz_i).$$

The estimation can be done easily by including $z_i$ and the product (interaction term) of $t$ and $z_i$ as two exposure variables. The VE curve is given by $VE(t) = 1 - HR(t) = 1 - exp(\beta_0 + \beta_1 t)$.

**Area under vaccine efficacy curve and number of cases averted**

It is useful to provide a summary efficacy measure and a public health impact measure, number of cases averted (NCA), where a case means an episode of the outcome event. We propose that VE during an interval of time from $t_1$ to $t_2$ can be quantified as the proportion (%) of area in this interval that lies below the VE curve, denoted $AUC_{t_1-t_2}$, where $t_1 = 0$ is typical. For simple VE curves, $AUC_{t_1-t_2}$ can be mathematically evaluated by plugging the Cox-type model estimates into an integral equation. For more complex VE curves, AUC can be evaluated by numerical integration functions available in major software like R. In our example, a mathematical solution is available:

$$AUC_{t_1-t_2} = \frac{\int_{t_1}^{t_2} 1 - exp(\hat{\beta}_0 + \hat{\beta}_1 t)dt}{t_2 - t_1} = \frac{\left[t_2 - \frac{1}{\hat{\beta}_1}exp(\hat{\beta}_0 + \hat{\beta}_1 t_2)\right] - \left[t_1 - \frac{1}{\hat{\beta}_1}exp(\hat{\beta}_0 + \hat{\beta}_1 t_1)\right]}{t_2 - t_1} \quad (3)$$

where $AUC_{t_1-t_2} = 1 - exp(\hat{\beta}_0)$ if $\hat{\beta}_1 = 0$.

In the literature, number of cases averted per 1000 person per unit of time has been calculated based on the empirically estimated step function of incidence rate over time[4,8,9]:

$$NCA_{SF} = 1000\left[\sum_{k=1}^{K}\left(\frac{E_{0k}}{T_{0k}} \cdot \Delta time_k\right) - \sum_{k=1}^{K}\left(\frac{E_{1k}}{T_{1k}} \cdot \Delta time_k\right)\right] \quad (4)$$

where $E_{0k}$ and $T_{0k}$ are respectively the number of outcome events and person-time in the control group in the $k$-th time interval ($k$=1, 2, 3, ..., $K$) since $t = 0$, $E_{1k}$ and $T_{1k}$ are their counterparts in the vaccine group, and $\Delta time_k$ is the duration of the $k$-th interval of follow-up time. For example, in the RTS,S/AS01 phase 3 trial, $\Delta time_k = 3$ months for all $k$. Note that if the trial duration is, e.g., $\sum_{k=1}^{K} \Delta time_k = 18$ months, equation (4) gives $NCA_{SF}$ per 1000 person-trial duration, meaning per 1500 person years. That "trial duration" was used as the time unit has not been always explicitly stated in the literature. Throughout the remainder of this article, we assume $\sum_{k=1}^{K} \Delta time_k = 12$ months for estimation of NCA unless otherwise stated.

$NCA_{SF}$ is subject to the influence of background disease incidence and the timing of vaccine delivery. We propose estimation of NCA using AUC and incidence data from the control arm:

$$NCA_{AUC} = 1000 \sum_{k=1}^{K} AUC_k \left(\frac{E_{0k}}{T_{0k}} \Delta time_k\right) \quad (5)$$

where $AUC_k$ is the area under VE curve during the $k$-th interval of time estimated by equation (3), and the rest are as defined previously for equation (4). To facilitate assessment of impact of launching vaccination in a new region that has a different disease incidence, we propose replacing $E_{0k}/T_{0k}$ in equation (5) by the recent disease incidence data from the target region.

If most trial participants completed vaccination ($t = 0$) around the beginning of low disease transmission season, as illustrated in Figure 1a, $NCA_{SF}$ would be small due to small $E_{0k}$ when $VE(t)$ is high but similar $E_{0k}$ and $E_{1k}$ when $VE(t)$ is low, or vice versa. Regardless of the actual timing of vaccination in a trial, we can estimate the NCA of seasonal vaccination strategy that times the completion of vaccination the to occur at the start of high season by adapting equation (5) to:

$$NCA_{AUC,season} = 1000 \sum_{k=1}^{K} AUC_k \left(\frac{E_{0,s+k-1}}{T_{0,s+k-1}} \Delta time_{s+k-1}\right) \quad (6)$$

where $s$ denotes the target month or quarter of completion of vaccination, $s + k$ denotes the $k$-th calendar month or quarter after $s$, and the sum over $K$ covers one year. For example, if completion of vaccination is targeted at July to optimize vaccine impact, then $s =$ July, $s + 1 =$ August and so on, and $K = 12$.

Conversely, an age-based vaccination strategy would approximate a situation that roughly the same number of people are vaccinated throughout time intervals in the year. The NCA under this delivery strategy can be approximated by applying equation (6) 12 times, each with a different $s$, and then taking the mean. This is denoted as $NCA_{AUC,age}$.

## Applications

### Simulation study

*Simulation methods*

We evaluate the performance of the proposed and conventional methods in a total of eight scenarios through simulation to mimic trials with different set-up but identical VE(t) = $1 - exp(\beta_0 + \beta_1 t)$. Many vaccines currently of public health importance concern repeatable outcome events such as episodes of malaria, pneumonia, RSV or rotavirus gastroenteritis. Therefore, we simulated repeatable event times and analyzed the simulated data by the AG model. Let each trial contain 1000 vaccinated and 1000 unvaccinated subjects. The planned study durations were set to be $\tau$ months, and the censoring times $C_i$ were either set to be $\tau$ for all $i$ or generated randomly and independently from $U(0.6\tau, \tau)$, i.e. uniform distribution function over the range $0.6\tau$ to $\tau$. We set the hazard function $\lambda_i(t) = \lambda_0(t)exp[(\beta_0 + \beta_1 t)z_i]$ where $\beta_0 = -4$ and $\beta_1 = 0.33$ for all trials in the simulation.

As shown in Table 1, the first four scenarios were simulated to evaluate the performance of methods under different planned follow-up times and with or without sample attrition, while maintaining the same baseline hazard of $\lambda_0(t) = 0.15$. The last four scenarios were simulated to assess performance of methods when vaccine delivery occurred in seasons with different baseline incidence rates and with or without sample attrition, using two step functions to replicate baseline hazards with seasonal variation. We set the baseline hazard for trials in which all participants completed vaccination at the beginning of the season with low transmission as

$$\lambda_{low}(t) = \begin{cases} 0.1 & 0 \leq t < 6 \\ 0.2 & 6 \leq t \leq 12 \end{cases}.$$

For trials that participants completed vaccination at the beginning of the season with high transmission, we set the baseline hazard to be

$$\lambda_{high}(t) = \begin{cases} 0.2 & 0 \leq t < 6 \\ 0.1 & 6 \leq t \leq 12 \end{cases}.$$

[Table 1 about here]

We applied the AG model with both PH assumption and time-varying intervention effect to estimate the parameters of interest. In the estimation of $NCA_{SF}$, $\Delta time$ was set to be 3 except that it was 1 for the last interval in scenarios of 10-month duration. The simulation used 1000 replicates. We used the thinning algorithm to simulate event times data[25]; details are provided in Online Supplementary Material 1.

*Simulation results*

The mean of 1000 estimates of regression parameters, VE, and AUC are summarized in Table 2. The mean of 1000 estimates of $NCA_{SF}$ and $NCA_{AUC}$ are summarized in Table 3. Results of the simulation verify the problems of the PH model and demonstrate the superiority of the model with time-varying effect. Although all scenarios have identical true VE pattern, the $\hat{\beta}$ estimated by PH model varied under different scenarios, and thus $\widehat{VE}$ also varied in the directions we discussed earlier. Specifically, PH model gave a lower $\widehat{VE} = 76.2\%$ for scenario 1 (12-month follow-up for all persons) and a higher $\widehat{VE} = 85.5\%$ for scenario 2 (10-month follow-up for all persons). PH model gave lower $\widehat{VE}$ for trials that started in the low transmission season than those started in the high season (scenarios 5 vs 6). Comparing the scenarios with sample attrition (scenarios 3, 4, 7 and 8) versus their counterparts with complete follow-up till 12 months (scenarios 1, 2, 5 and 6, respectively), there was a consistent pattern that $\widehat{VE}$ tended to be higher in the presence of sample attrition.

[Table 2 about here]

On the other hand, $\hat{\beta}_0$ and $\hat{\beta}_1$ estimated by the model with time-varying intervention effect are consistent for all eight scenarios despite variations in trial features (Table 2). Furthermore, $AUC_{0-\tau}$ estimated by the model with time-varying effect are almost identical to the $\widehat{VE}$ estimated by PH model when there was no sample attrition and seasonality regardless of planned trial duration [scenarios 1 ($\tau = 12$) and 2 ($\tau = 10$)], demonstrating the equivalence between VE and $AUC_{0-\tau}$ in this situation. Contrary to PH model, the model with time-varying effects gave practically the same $AUC_{0-\tau}$ with or without sample attrition. Moreover, $AUC_{0-\tau}$ is not sensitive to timing of vaccine delivery.

VE curves from trials with 12-month duration can be used to accurately estimate $AUC_{0-10}$ by integrating the area under curve from 0 to 10 months, and vice versa. The last two columns of Table 2 demonstrate this feature of the method.

Table 3 shows estimated $NCA_{SF}$ and $NCA_{AUC}$ per 1000 persons up to 10 months and 12 months. In scenarios without sample attrition (1, 2, 5 and 6), $NCA_{AUC}$ is practically identical to $NCA_{SF}$. In the other scenarios that involved sample attrition, $NCA_{AUC}$ is consistent with the corresponding $NCA_{AUC}$ in scenarios without sample attrition, while $NCA_{SF}$ tend to increase especially when evaluated over 12 months. Moreover, $NCA_{AUC,season}$ estimated in scenarios 5 and 7 (trials begin in low season) agreed with $NCA_{AUC}$ in scenarios 6 and 8 (trials begin in high season). Comparison of $NCA_{AUC,season}$ and $NCA_{AUC,age}$ demonstrates the importance of delivery strategy given disease seasonality.

[Table 3 about here]

## Demonstration with data from phase 3 trial of RTS,S malaria vaccine

*Trial design and analytic methods*

Data from a phase 3 randomized controlled trial of RTS,S/AS01 malaria vaccine were re-analyzed. The trial recruitment was conducted across 11 African sites between March 2009 and January 2011[26]. Children aged 5 to 17 months and young infants aged 6 to 12 weeks were separately randomized (1:1:1) to receive three doses of RTS,S/AS01 at months 0, 1, and 2 and booster dose at month 20 (R3R group), or three doses of RTS,S/AS01 at months 0, 1, and 2 and comparator vaccine at month 20 (R3C group), or a comparator at months 0, 1, and 2 and 20 (C3C group).

For illustration, we used data of children to estimate efficacy of the primary vaccination during the 17.5-month duration from 14 days after dose 3 to before the booster dose, and therefore R3R group and R3C group were combined as the intervention arm. The per-protocol population included 4589 and 2341 children who were allocated to the intervention and control arm, respectively. Clinical malaria episodes (primary case definition) were as defined in the original RTS,S/AS01 trial.

To mimic scenarios of different trial durations, the analysis time was truncated to 14.5 months and 11.5 months in addition to the analysis of 17.5 months. AG models with PH assumption and with time-varying effect were applied, with stratification on sites. The RTS,S trial team previously found that the time-varying effect functions $\beta_0 + \beta_1 ln(t)$ and $\beta_0 + \beta_1 \sqrt{t}$ each fitted the data from some but not all study sites well[26]. The $ln(t)$ function fitted better than the $\sqrt{t}$ function for Nanoro, Burkina Faso, a site that has the strongest seasonality of malaria transmission[26], which we will pay particular attention to. In this section we illustrate the methods using $\beta_0 + \beta_1 ln(t)$ as the time-varying effect function. In Online Supplementary Material 2 we provide sensitivity analysis including the $\sqrt{t}$ function.

Malaria disease incidence in Burkina Faso exhibits a sharp seasonal upsurge in June and subsiding after December[27]. For the study site in Burkina Faso, $NCA_{SF}$ and $NCA_{AUC}$ were estimated using equations 4 and 5, respectively, both with $\Delta time$ =1 month and $T = 12$ months starting from 14 days after dose 3. We also estimated $NCA_{AUC,season}$ and $NCA_{AUC,age}$, using data from the control group at this site from June 2010 to May 2011 for $E_{0,s+k-1}$ and $T_{0,s+k-1}$ in equation 6).

Follow-up to end of the 17.5-month duration was almost complete across all sites (93.2%) and at Nanoro, Burkina Faso (92.2%), so we do not expect sample attrition to generate any substantive influence on the findings.

*Findings*

Analysis results across all sites were summarized in Table 4. Using the AG model with PH assumption, $\widehat{VE}$ estimates (95% CI) for trial durations of 17.5, 14.5 and 11.5 months are 42.3% (38.2% to 46.2%), 45.8% (41.8% to 49.5%), and 49.6% (45.8% to 53.1%), respectively, demonstrating the expected variation of higher $\widehat{VE}$ for shorter trial duration. On the other hand, $\hat{\beta}_0$, $\hat{\beta}_1$, $AUC_{0-17.5}$, $AUC_{0-14.5}$, and $AUC_{0-11.5}$ estimated by the model with time-varying intervention effect based on three different analysis durations are approximately the same and the AUC estimates are within the 95% CI of the $\widehat{VE}$ estimated from data of the same duration.

[Table 4 about here]

For data of Nanoro, Burkina Faso alone in the 17.5-month follow-up period, the PH model gave $\widehat{VE} = 42.0\%$ (95% CI 34.1% to 48.9%). The model with time-varying intervention effect gave $\widehat{VE}(t) = 1 - \exp[-1.66 + 0.525 \ln(t)]$ and $AUC_{0-17.5} = 43.7\%$. Additionally, $NCA_{SF} = 1362$ and $NCA_{AUC} = 1434$. Further examination of the disease incidence data by months shows that the difference between $NCA_{SF}$ and $NCA_{AUC}$ was mainly due to a very small incidence rate difference (IRD) in the 8[th] month, which was much smaller than the IRD's in the two preceding and two subsequent months (details in Online Supplementary Materials 3). $NCA_{SF}$ captured this idiosyncratic feature of the data while $NCA_{AUC}$ smoothed this out.

In the Nanoro, Burkina Faso study site, children's date of completion of vaccination ranged from December 2009 to May 2010, with median in February 2010, approximately the middle of the low malaria transmission season. This would likely under-estimate the true impact of RTS,S vaccination as in real-life such timing of vaccination is unlikely. Figure 2 shows the estimates of $NCA_{AUC}$ per 1000 person-year that targets different calendar months as the first month post completion of vaccination. Under a seasonal vaccination strategy that aimed at completion in July, $NCA_{AUC,season} = 2459$. The mean of the 12 values in Figure 2 gives $NCA_{AUC,age} = 1920$.

[Figure 2 about here]

## Discussion

A recent study of malaria chemoprevention proposed using area under protective efficacy curve as a summary measure of intervention effect[15], but it did not consider whether the approach may mitigate the influences of trial implementation features. In this article, we proposed and demonstrated using simulation and real data that AUC of the VE trajectory is a summary measure that remains robust despite variations in trial implementation features.

It is not widely appreciated that VE estimates based on an invalid PH assumption are affected by magnitude of sample attrition, which was clearly demonstrated by our simulation study. In real-life, people continuously experience health outcomes with or without being placed under observation. Conventional estimates of VE based on PH model and data from trials with substantial sample attrition risk do not reflect real-life. The proposed method does not suffer the bias arising from sample attrition.

Clearly defining the estimand corresponding to the clinical question of interest facilitates trial design, estimation and interpretation that fits the purpose[28]. For example, the estimand of the R21 phase 2 malaria vaccine trial was VE over 12-month if the vaccine was deployed seasonally to achieve maximum impact[29]. The use of our methods to estimate $AUC_{0-\tau}$ and various NCA measures has the potential to improve comparison and evidence synthesis using data from trials designed for different estimands.

An alternative way to present vaccine impact is by Number Needed to Vaccinate to prevent one case (NNV)[12,30]. Given that in equation (5) $NCA_{AUC}$ is defined per 1000 persons for a time unit,

$$NNV = 1000/NCA_{AUC}$$

In studies of repeatable events such as malaria disease episodes, a person may become a case multiple times and $NNV$ may be smaller than one. In that case Number Need to Vaccinate to prevent 1000 cases may be more appropriate. Estimation results of $NCA_{AUC} = 1434$, $NCA_{AUC,season}=2459$, and $NCA_{AUC,age} = 1920$ from RTS,S/AS01 trial data can be translated into NNV, indicating approximately 697, 407, and 521 subjects need to be vaccinated to prevent 1000 malaria cases in one year if vaccines were delivered as in the Burkina Faso site of the RTS,S/AS01 trial, a seasonal strategy, or by age, respectively. These estimates allow us to easily interpret impact of the intervention and suggest that the seasonal strategy is the more impactful approach to deliver vaccines in regions with highly seasonal transmission.

Other challenges in synthesizing trial results include differences in study protocols, such as variations in definition of primary endpoints, the types of placebos used, and the characteristics of study populations[30]. This article has not discussed these issues. Moreover, it is not guaranteed that a vaccine demonstrating a certain pattern of VE(t) in one study population will exhibit the same pattern in another population. Our proposal of plugging past disease incidence data from a new region expecting to implement vaccination into equations (5) or (6) to gauge vaccine impact in the new region should be interpreted with this uncertainty in view. In this study, we focused on the concept and applications of VE(t) and AUC. A limitation of this work is that we have not discussed how best to estimate VE(t) trajectory. Further research may explore whether there exists flexible yet parsimonious time functions that broadly perform well in vaccine research.

**Acknowledgments**

We thank www.clinicalstudydatarequest.com and GlaxoSmithKline for providing access to the anonymized RTS,S/AS01 trial data.

**Funding**

This work was supported by the Duke-NUS Signature Research Programme funded by the Ministry of Health, Singapore.

**Ethics approval**

Analysis of the de-identified data was approved by National University of Singapore Institutional Review Board (ref: NUS-IRB-2022-510).

**Conflict of interest**

The authors declare that they have no conflict of interest with respect to this research study and paper.

**Contributors**

ZZ conducted simulation and data analysis and wrote the draft and final version of the manuscript. XM participated in analysis programming, data management and interpretation of findings, and critically reviewed and revised the draft version of the article. PM participated in the conceptualization and interpretation of findings, and critically reviewed and revised the draft version of the article. YBC conceived the study, designed the study, interpreted the findings, and critically reviewed and revised the draft version of the article. All authors approved the final version and agreed to be accountable for all aspects of the work.

**Data availability statement**

The RTS,S/AS01 trial data is owned by GlaxoSmithKline. The authors obtained access to the data (#GSK-110021) via application to and permission by www.clinicalstudydatarequest.com. Request for access to the data may be submitted to www.clinicalstudydatarequest.com for the data owner's consideration.

Table 1. Simulation scenarios varied by planned study durations, sample attrition, and baseline incidence patterns.

| Scenario | Features | $\tau$ | $C_i$ | $\lambda_0(t)$ |
|---|---|---|---|---|
| 1 | long, without attrition | 12 | 12 | 0.15 |
| 2 | short, without attrition | 10 | 10 | 0.15 |
| 3 | long, with attrition | 12 | $U(7.2,12)$ | 0.15 |
| 4 | short, with attrition | 10 | $U(6,10)$ | 0.15 |
| 5 | low-high, without attrition | 12 | 12 | $\lambda_{low}(t)$ |
| 6 | high-low, without attrition | 12 | 12 | $\lambda_{high}(t)$ |
| 7 | low-high, with attrition | 12 | $U(7.2,12)$ | $\lambda_{low}(t)$ |
| 8 | high-low, with attrition | 12 | $U(7.2,12)$ | $\lambda_{high}(t)$ |

Table 2. Simulation results of parameter, VE, and AUC estimation based on AG model with PH assumption and time-varying intervention effect in eight scenarios.

| Scenario | Proportional Hazard | | Time-varying Intervention Effect | | | |
|---|---|---|---|---|---|---|
| | $\hat{\beta}$ | $\widehat{VE}$ (%) | $\hat{\beta}_0$ | $\hat{\beta}_1$ | $AUC_{0-12}$ (%) | $AUC_{0-10}$ (%) |
| 1 | -1.44 | 76.2 | -4.01 | 0.331 | 76.1 | 85.5[a] |
| 2 | -1.93 | 85.5 | -4.01 | 0.330 | 76.1[b] | 85.5 |
| 3 | -1.93 | 85.4 | -4.02 | 0.331 | 76.2 | 85.6[a] |
| 4 | -2.33 | 90.2 | -4.01 | 0.330 | 75.7[b] | 85.4 |
| 5 | -1.21 | 70.2 | -4.02 | 0.332 | 76.2 | 85.5 |
| 6 | -1.73 | 82.2 | -4.01 | 0.331 | 76.1 | 85.5 |
| 7 | -1.68 | 81.3 | -4.01 | 0.331 | 76.0 | 85.4 |
| 8 | -2.19 | 88.8 | -4.01 | 0.331 | 76.0 | 85.5 |

a.     $AUC_{0-10}$ for scenario 1 and 3 are interpolated from data with study duration up to 12 months.

b.     $AUC_{0-12}$ for scenario 2 and 4 are extrapolated from data with study duration up to 10 months.

Table 3. Simulation results of $NCA_{SF}$ and $NCA_{AUC}$ per 1000 persons up to 10 months and 12 months.

| Scenario | $NCA_{SF}(12)$[a] | $NCA_{SF}(10)$ | $NCA_{AUC}(12)$[a] | $NCA_{AUC}(10)$ | $NCA_{AUC,season}(12)$[b] | $NCA_{AUC,age}(12)$[b] |
|---|---|---|---|---|---|---|
| 1 | 1370.3 | 1281.6 | 1370.2 | 1281.7 | | |
| 2 | | 1282.4 | | 1282.4 | | |
| 3 | 1422.9 | 1287.8 | 1372.9 | 1283.5 | | |
| 4 | | 1294.8 | | 1281.7 | | |
| 5 | 1263.1 | 1144.2 | 1262.7 | 1144.7 | 1479.3 | 1360.2 |
| 6 | 1480.0 | 1420.8 | 1479.7 | 1420.8 | 1479.7 | 1359.9 |
| 7 | 1326.4 | 1149.0 | 1259.4 | 1142.7 | 1477.1 | 1357.0 |
| 8 | 1512.9 | 1425.0 | 1480.5 | 1421.7 | 1478.7 | 1359.0 |

a.     $NCA_{SF}$ and $NCA_{AUC}$ up to 12 months are unavailable for scenario 2 and 4 as outcome incidence data was simulated only up to 10 months.

b.     $NCA_{AUC,season}$ and $NCA_{AUC,age}$ up to 12 months for scenario 1 to 4 are identical to $NCA_{AUC}$ up to 12 months as there was no seasonality in the outcome incidence.

Table 4. Estimations of vaccine efficacy from RTS,S/AS01 trial data using different methods and follow-up time for analysis.

| Follow-up (month) | $\hat{\beta}$ | $\widehat{VE}$ (%) | $\hat{\beta}_0$ | $\hat{\beta}_1$ | $AUC_{0-17.5}$ (%) | $AUC_{0-14.5}$ (%) | $AUC_{0-11.5}$ (%) |
|---|---|---|---|---|---|---|---|
| 17.5 | -0.550 | 42.3 | -1.349 | 0.392 | 42.7 | 46.8 | 51.4 |
| 14.5 | -0.612 | 45.8 | -1.363 | 0.401 | 42.4 | 46.6 | 51.3 |
| 11.5 | -0.685 | 49.6 | -1.354 | 0.393 | 42.9 | 46.9 | 51.6 |

Figure 1. Wanning vaccine efficacy over time plotted on different seasonal patterns of disease incidence.

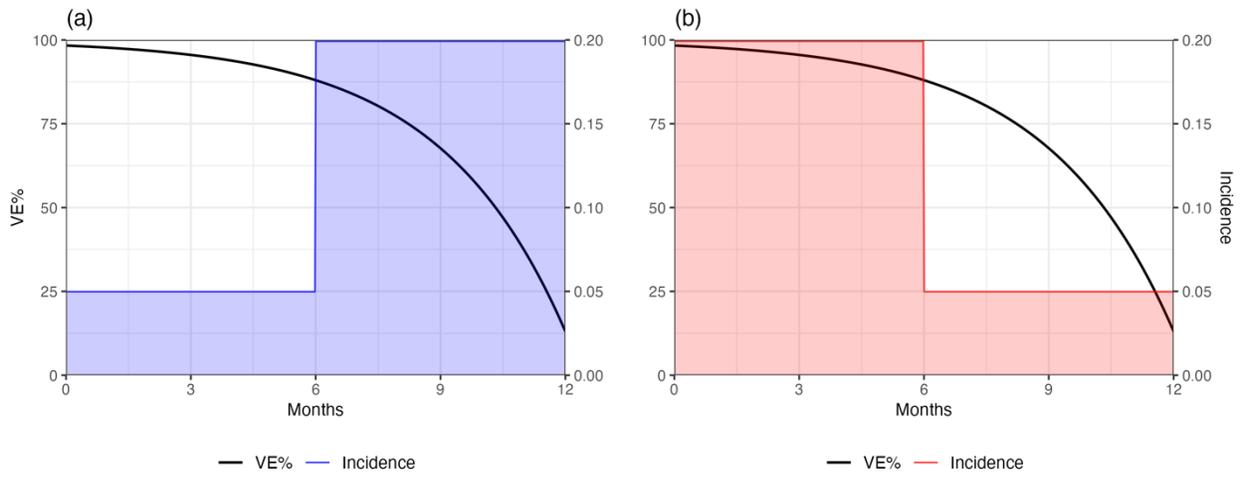

Figure 2. $NCA_{AUC}$ estimates using equation (6) targeting completion of vaccination in different calendar months.

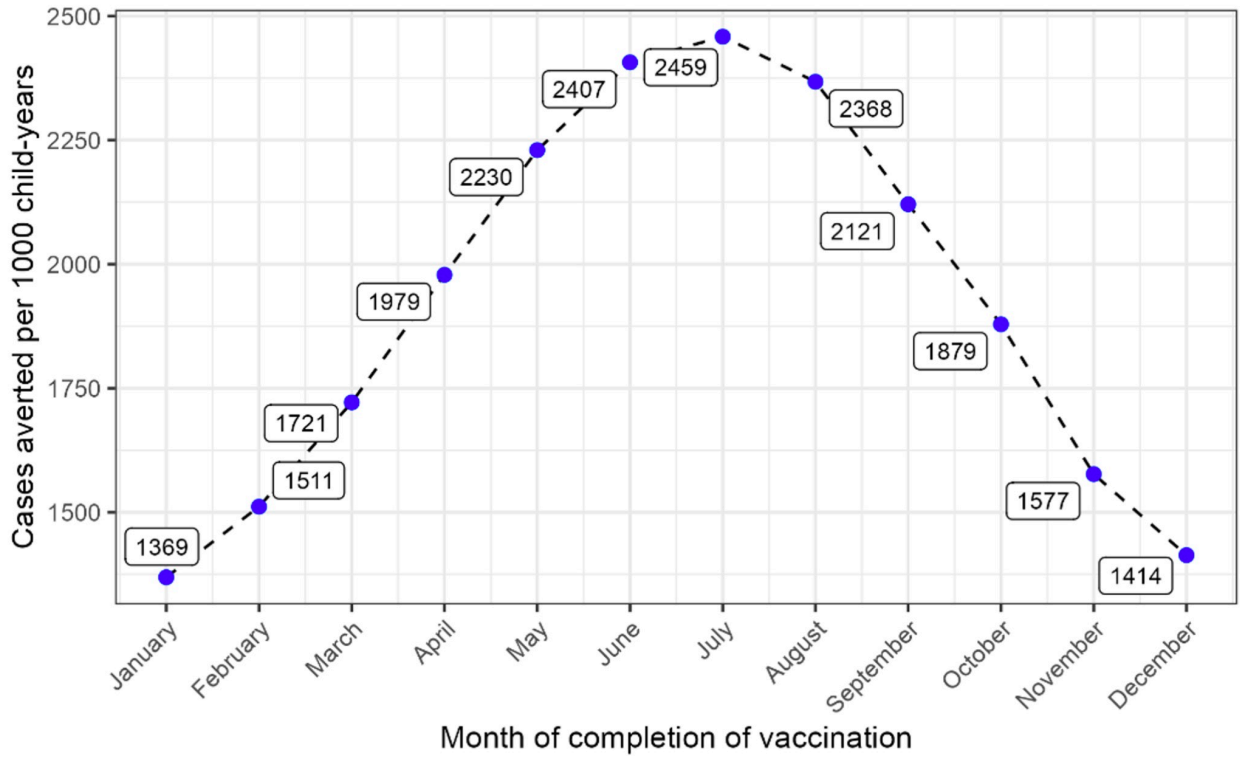

# Online Supplementary Material 1

*Algorithm for recurrent event times simulation*

For the $i$th subject in each dataset, the $j$th recurrent event time $T_{ij}$ was generated according to the hazard function $\lambda_i(t)$ using the thinning approach with the following steps (Lewis & Shedler, 1979):

1. Choose a value $\bar{\lambda}$ such that $\forall t, \lambda_i(t) \leq \bar{\lambda}$. Set $T_{i0} = 0, T^* = 0, j = 1$.

2. Draw a random number $R \sim Exp(\bar{\lambda})$ and update $T^* = T^* + R$.

3. When $T^* \leq C_i$, generate $V \sim U(0,1)$.

   – If $V \leq \frac{\lambda(T^*)}{\bar{\lambda}}$, then set $T_{ij} = T^*$ and censoring status $\delta_{ij} = 1$, update $j = j + 1$, and return to step 2.

   – If $V > \frac{\lambda(T^*)}{\bar{\lambda}}$, then return to step 2 without updating $T_{ij}$ and $j$.

   When $T^* > C_i$, stop the process and set censoring status $\delta_{ij} = 0$ at $C_i$.

**Reference**

Lewis PA, Shedler GS. Simulation of nonhomogeneous Poisson processes by thinning. *Naval Research Logistics Quarterly* 1979; 26(3): 403–413.

# Online Supplementary Material 2

*Sensitivity analysis of VE curve and AUC estimation using RTS,S malaria vaccine data*

Using per protocol analysis data set of 17.5 months' follow-up duration, the BIC value for the AG models with time-varying effect function $f(t) = \beta_0 + \beta_1 ln(t)$ was 4.3 larger than that for the AG model using function $f(t) = \beta_0 + \beta_1 \sqrt{t}$. Using the BIC values to approximate Bayes Factor give BF=8.6, which is "positive" but not "strong" evidence in favor of $\sqrt{t}$ (Wagenmakers, 1995).

Figure S1 shows the fitted VE curves using the two functions. Using $ln(t)$ function, the VE curves based on 17.5-, 14.5- and 11.5-month data almost completely overlap. The VE curve estimation is not sensitive to the data duration. The curves show that the decline of VE was decelerating.

Using $\sqrt{t}$ function, the decline of VE did not appear to decelerate. Moreover, the VE curves based on different duration of data visibly differ, with shallower slope for VE curves based on longer duration of data.

Figure S1. Fitted VE curves.

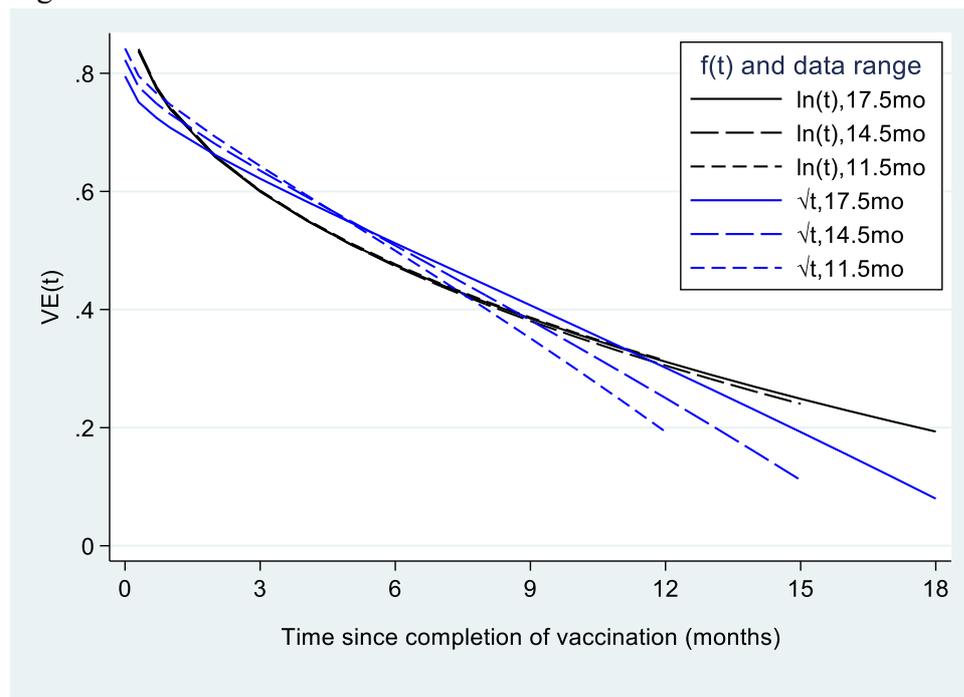

Table S1 shows the estimation of VE, VE curves and $AUC$. Regardless of the choice of $\ln(t)$ or $\sqrt{t}$ function, the $AUC$ and $\widehat{VE}$ estimated from the same duration of data never differ by more than 2%. The biggest difference was between $\widehat{VE} = 49.6\%$ in the analysis based on 11.5 months' data and $AUC_{0-11.5} = 51.6\%$ estimated using $\ln(t)$ function. Furthermore, the $AUC$ estimated by $\ln(t)$ or $\sqrt{t}$ function using the same duration of data never differ by more than 1%. The biggest difference was between $AUC_{0-14.5} = 46.6\%$ based on $\ln(t)$ function and $AUC_{0-14.5} = 45.6\%$ based on $\sqrt{t}$ function, both based on analysis of 14.5 months' data.

Table S1. Estimates of vaccine efficacy from RTS,S data using different methods

| f(t) | Duration (month) | $\hat{\beta}$ | $\widehat{VE}$(%) | $\hat{\beta}_0$ | $\hat{\beta}_1$ | $AUC_{0-17.5}$(%) | $AUC_{0-14.5}$(%) | $AUC_{0-11.5}$(%) |
|---|---|---|---|---|---|---|---|---|
| ln(t) | 17.5 | -0.550 | 42.3 | -1.349 | 0.392 | 42.7 | 46.8 | 51.4 |
| ln(t) | 14.5 | -0.612 | 45.8 | -1.363 | 0.401 | 42.4 | 46.6 | 51.3 |
| ln(t) | 11.5 | -0.685 | 49.6 | -1.354 | 0.393 | 42.9 | 46.9 | 51.6 |
| $\sqrt{t}$ | 17.5 | -0.550 | 42.3 | -1.589 | 0.354 | 42.0 | 47.3 | 52.7 |
| $\sqrt{t}$ | 14.5 | -0.612 | 45.8 | -1.734 | 0.417 | 38.8 | 45.6 | 52.2 |
| $\sqrt{t}$ | 11.5 | -0.685 | 49.6 | -1.850 | 0.471 | 35.3 | 43.4 | 51.1 |

Fitting the ln(t) function to the data of longer duration (e.g. 17.5 months) and then using the estimated VE curve to obtain $AUC$ for shorter duration (e.g. 14.5 months) gives good prediction of $\widehat{VE}$ estimated from shorter duration of data. For example, based on this ln(t) function and 17.5 months' of data, the VE curve can give interpolation of $AUC_{0-14.5} = 46.8\%$, which agrees well with $\widehat{VE} = 45.8\%$ estimated using PH model and 14.5 months' data. The converse is also true, i.e. extrapolation based on VE curve estimated from shorter duration to obtain AUC for longer duration gives AUC estimates that agree well with the $\widehat{VE}$ estimated from the longer duration of data.

Fitting the $\sqrt{t}$ function to data of longer duration and then using the estimated VE curve used to interpolate $AUC$ for shorter duration gives good prediction of $\widehat{VE}$ estimated from shorter duration of data. However, the converse is not true. For example, using 11.5 month's data to estimate VE curve and then extrapolate to 17.5 months gives $AUC_{0-17.5} = 35.3\%$, as opposed to $\widehat{VE} = 42.3\%$ estimated from PH model using 17.5 month's data, differing by 7%.

In conclusion, estimation of AUC as a summary measure of efficacy is not sensitive to the choice of f(t). Using VE curves estimated from trials of longer duration to interpolate AUC for shorter duration for comparison with $\widehat{VE}$ estimated from trials of shorter duration is not sensitive to the choice of f(t) either. However, the converse is not true. Furthermore, the ln(t) function appears to be less sensitive to trial duration than $\sqrt{t}$ function if waning is decelerating.

**Reference**
Wagenmakers EJ. A practical solution to the pervasive problems of *p* values. *Psychonomic Bulletin Review* 2007; 14(5): 779-804.

# Online Supplementary Material 3

*$NCA_{SF}$ and $NCA_{AUC}$ estimated from data of Nanoro, Burkina Faso*

For the study site in Burkina Faso, $NCA_{SF}$ and $NCA_{AUC}$ were estimated using equations 4 and 5 with $\Delta time$ = 1 month and $T$ = 12 months. The conventional estimate of $NCA_{SF} = 1362$ is around 5% lower than $NCA_{AUC} = 1434$. We further examine incidence of the control arm (i.e. $E_{0k}/T_{0k}$), incidence of the intervention arm (i.e. $E_{1k}/T_{1k}$), incidence rate difference (IRD) $AUC_k$ for each of the $k$-th time interval ($k$=1, 2, 3, …, 12) derived from the time-varying effect model, and estimates of monthly $NCA_{SF}$ and $NCA_{AUC}$.

Table S2. Monthly estimates of $NCA_{SF}$ and $NCA_{AUC}$ in the first 12 months since completion of vaccination, Burkina Faso.

| $k$ | $E_{1k}/T_{1k}$ | $E_{0k}/T_{0k}$ | $IRD_k$ [a] | $AUC_k$ | $NCA_{SF,k}$ | $NCA_{AUC,k}$ | $NCA_{AUC,k} - NCA_{SF,k}$ |
|---|---|---|---|---|---|---|---|
| 1 | 7.7 | 23.4 | 15.6 | 0.875 | 16 | 20 | 5 |
| 2 | 0.0 | 10.2 | 10.2 | 0.765 | 10 | 8 | -2 |
| 3 | 5.2 | 5.1 | -0.1 | 0.692 | 0 | 4 | 4 |
| 4 | 7.9 | 51.9 | 44.0 | 0.632 | 44 | 33 | -11 |
| 5 | 48.4 | 192.3 | 143.9 | 0.580 | 144 | 111 | -32 |
| 6 | 140.2 | 389.6 | 249.4 | 0.533 | 249 | 208 | -42 |
| 7 | 312.7 | 666.7 | 354.0 | 0.490 | 354 | 327 | -27 |
| 8 | 458.5 | 563.5 | 105.0 | 0.450 | 105 | 254 | 149 |
| 9 | 375.7 | 547.5 | 171.8 | 0.413 | 172 | 226 | 54 |
| 10 | 215.0 | 377.1 | 162.1 | 0.377 | 162 | 142 | -20 |
| 11 | 136.4 | 213.8 | 77.4 | 0.344 | 77 | 73 | -4 |
| 12 | 66.9 | 98.8 | 31.9 | 0.312 | 32 | 31 | -1 |
| | | | | Total:[b] | 1365 | 1437 | 71 |

a. $IRD_k = E_{0k}/T_{0k} - E_{1k}/T_{1k}$.
$NCA_{SF,k} = (E_{0k}/T_{0k} - E_{1k}/T_{1k}) \times 1$.
$NCA_{AUC,k} = AUC_k \times E_{0k}/T_{0k} \times 1$.
b. These values are slightly different from values shown in main text due to rounding.

Table S2 shows that the difference between the total $NCA_{SF}$ and $NCA_{AUC}$ is mainly due to the small IRD in the 8[th] month (i.e. $k$=8). This IRD is substantially smaller than the IRD's in the two preceding months and two subsequent months. $NCA_{SF}$ captured this idiosyncratic feature of the data while $NCA_{AUC}$ smoothed this out. Therefore, the latter gives a higher *NCA* for this month, which is a major contributor to the observation of $NCA_{SF}$ smaller than $NCA_{AUC}$ over the 12-month duration. While we consider the $NCA_{SF}$ estimate an accurate quantification of vaccine impact for this specific dataset, the $NCA_{AUC}$ estimate is likely more suitable for making inference of the vaccine's impact in similar trials of the vaccine conducted under similar conditions.